\def\year{2019}\relax
\documentclass[letterpaper]{article} 
\usepackage{aaai19}  
\usepackage{times}  
\usepackage{helvet} 
\usepackage{courier}  
\usepackage[hyphens]{url}  
\usepackage{graphicx} 
\usepackage{graphicx}  
\usepackage{tabulary}
\usepackage{comment}

\nocopyright
\title{Assessing Regulatory Risk in Personal Financial Advice Documents: a Pilot Study}
\author{Wanita Sherchan,
Simon Harris,
Sue Ann Chen, 
Nebula Alam, \\ \Large \textbf{
Khoi-Nguyen Tran,
Adam J. Makarucha, 
Christopher J. Butler}\\
IBM Research Australia  \\
\{wanita.sherchan, siharris, sachen, anebula, khndtran, adamjm, chris.butler\}@au1.ibm.com \\
}
\begin{document}

\maketitle

\begin{abstract}

Assessing regulatory compliance of personal financial advice is currently a complex manual process. In Australia, only $5\%$-$15\%$ of advice documents are audited annually and $75\%$  of these are found to be non-compliant\cite{ASICRep562}. This paper describes a pilot with an Australian government regulation agency where Artificial Intelligence (AI) models based on techniques such natural language processing (NLP), machine learning and deep learning were developed to methodically characterise the regulatory risk status of personal financial advice documents. The solution provides traffic light rating of advice documents for various risk factors enabling comprehensive coverage of documents in the review and allowing rapid identification of documents that are at high risk of non-compliance with government regulations. This pilot serves as a case study of public-private partnership in developing AI systems for government and public sector.

\end{abstract}

\section{Introduction}

Demand for financial advisory services has been increasing rapidly in recent years. A global average of $46\%$ citizens per country sought professional financial advice in 2017 \cite{Ins201708} . In Australia, financial advisors need to abide by government regulatory requirements to act in the ``best interests'' of the client \cite{FOFA2012,ASICguidelines}.
To ensure compliance with these regulations, a small percentage of financial advice files are audited by the Australian government each year. These manual audits are complex, time consuming (taking between 2 to 6 hours per document) and applied to only a small sample - covering around $5\%$ to$15\%$ of the Statement of Advice (SoA) documents produced) \cite{ASICRep515}. Advisors also often have autonomy in choosing documents for audit. Even with biased sampling, Australia's regulatory agency the Australian Securities and Investments Commission (ASIC) revealed that $75\%$ of the advice documents reviewed were non-compliant \cite{ASICRep562}. Furthermore, often the review/audit occurs years after the advice has been executed, meaning a large portion of Australian citizens have already suffered significant financial losses which the subsequent audits would not be able to prevent.

In late 2017, an inquiry was established by the Australian Government to report on misconduct in the financial services industry, known as The Royal Commission into Misconduct in the Banking, Superannuation and Financial Services Industry \cite{AusRoyalCommission2019}. The Royal Commission report was published in 2019 and included a list of 76 recommendations to ensure Australian citizens receive financial advice that is truly in their ``best interest''. Many recommendations tighten the governance procedures for financial institutions within Australia and consequently place additional burden of ensuring regulatory compliance on government regulation agencies.

To address these issues we have developed a system that leverages NLP, Machine Learning (ML) and Deep Learning (DL) techniques to ingest SoA documents and apply traffic light ratings to highlight various risk factors within an SoA. Our system can be used by financial advisors while they are in the process of creating the SoAs to assess advice compliance before it is delivered to the clients. In this manner clients are protected from suffering unnecessary financial loss, whilst advisors are protected from violating regulatory compliance requirements. The system can also be used by government regulatory compliance auditors to enable rapid review of a large corpus of SoAs, highlighting those at greatest risk of non-compliance.

We conducted a pilot of our system with an Australian government regulation agency. This paper describes the pilot study, challenges, outcomes and lessons learned in developing AI techniques in cooperation with a government agency with the aim of benefiting its citizens.

\section{Pilot Study }

The pilot study was conducted with an Australian Government agency focused on enforcing and regulating company and financial services laws to protect Australian consumers, investors and creditors. The agency is responsible for monitoring the quality of financial advice provided to citizens. Poor financial advice can undermine investor and consumer trust and confidence in the financial system. To test the quality of financial advice provided to consumers, each year, the agency reviews a significant number of client files containing records of financial advice (`client file reviews'). These reviews are typically conducted on a sampling basis by manually referencing the applicable statutory laws and advisor conduct obligations. The main objective of the pilot study was to apply AI (NLP, ML, DL) technologies to aid the file reviews process through a two-fold approach - map statutory advice conduct obligations and relevant sections of the law to a number of risk indicators, and apply AI techniques to rapidly identify advice files that are at high risk of being non-compliant.

The pilot was conducted from July 2018 to September 2018.

\subsection{Pilot Study Design }

The regulatory agency typically uses a number of client files as part of the review process:

\begin{itemize}
\item `Statement of Advice (SoA)': a legal document between a client and a financial advisor that lists the client's goals and objectives, the advisor's recommendations for achieving those goals and objectives and the reasoning behind the recommendations;
\item `Fact Finder': a document recording personal circumstances of the client that the advisor has obtained from discussions with the client;
\item file notes recording discussions with the client;
\item research demonstrating the advisor's investigation of the subject matter of the advice;
\item Product Disclosure Statements (PDS) for the products that have been recommended;
\item product application forms; and
\item working papers
\end{itemize}

Out of these documents, SoA documents are mandatory for every client file where personal advice has been provided.
Therefore, the pilot focused on developing a set of Key Risk Indicators (KRIs) that could be extracted from the SoA documents. Further, over the preceding years, the agency had found financial advice covering Self Managed Superannuation Funds (SMSFs) to be the most non-compliant with respect to various regulations. Therefore, the pilot focused on KRIs that are most relevant to the SMSF domain and subsequently a sample of SoA documents from this domain were selected for the pilot.

The government agency was keen to understand the process required to create AI models and validate that models in this domain could be created and evaluated within a short period of time. As such, an aggressive target was set to create and validate 6 KRI models during the 5 week implementation period.

The pilot was structured in two phases. Phase 1 (design phase) consisted of a series of workshops in which IBM Researchers worked alongside government Subject Matter Experts (SMEs) to (i) define general and SMSF specific key risk indicators of non-compliance, (ii) prioritise the risk indicators based on their impact and the potential of the indicator to be extracted from the SoA documents and (iii) determine the methodologies to extract these indicators. Phase 2 (implementation phase) consisted of the technical implementation of the methodologies on a set of sample documents, testing and presentation of the findings.

\subsection{Methodology for Determining the Key Risk Indicators (KRIs) }

\begin{figure*}[hbt!]
\centering
{\includegraphics[width=0.8\textwidth]{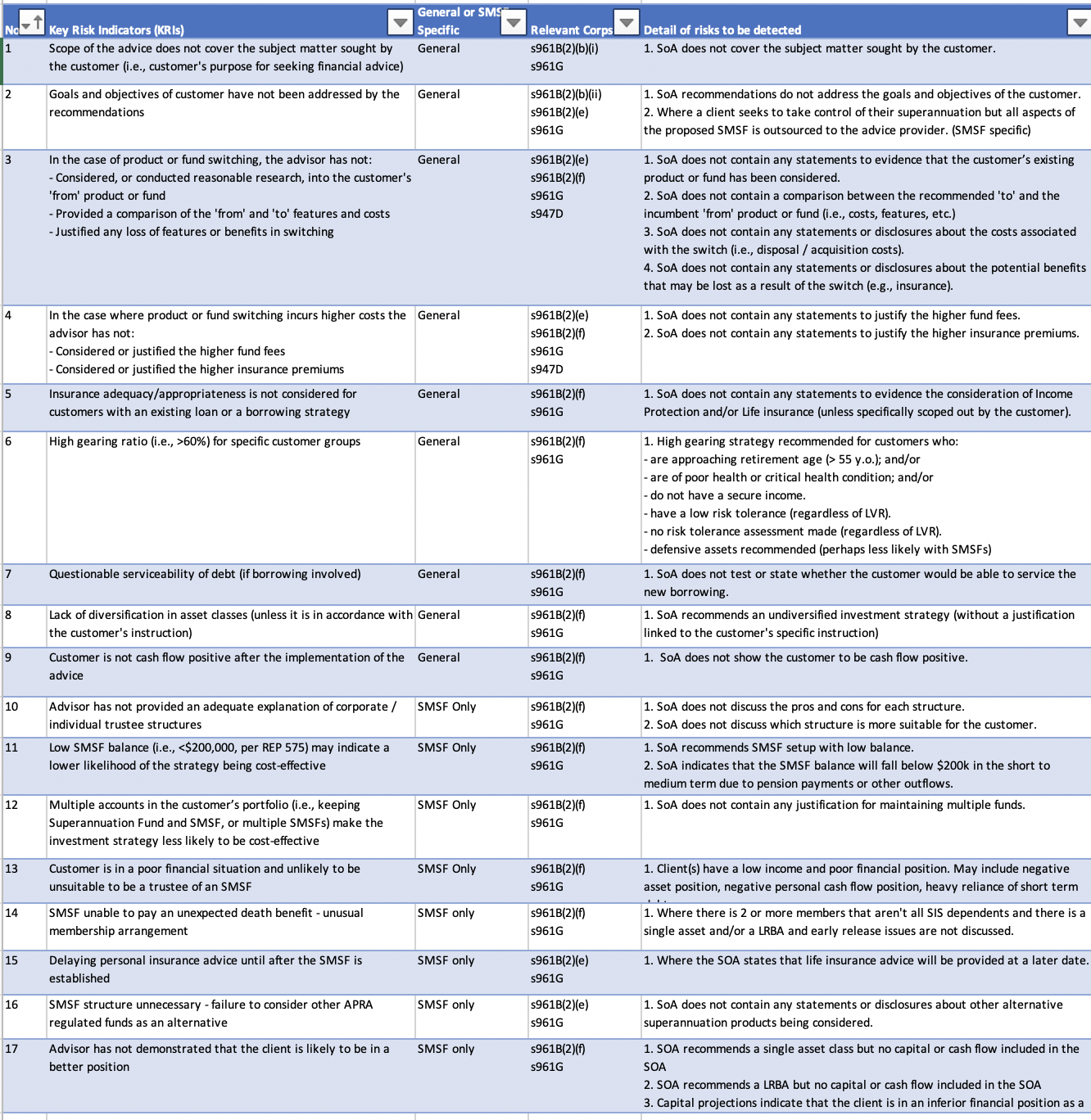}}
\caption{Full list of Key Risk Indicators (KRIs) of non-compliance in personal financial advice.}
\label{full-kris}
\end{figure*}

The first part of the pilot was focused on translating statutory advice conduct obligations and relevant sections of the law to a number of risk indicators and defining the details of the risks to be detected. For this part, researchers worked with auditors and analysts from the regulatory agency and consultants from Promontory Financial Group\footnote{https://www.promontory.com}, a global consulting firm focused on risk management and regulatory compliance in financial services sector, who was a partner in this pilot.

For determining the indicators of risk, the team started with mapping and interpreting the relevant sections of the law and statutory advice conduct obligations related to personal financial advice to come up with 17 indicators of risk as shown in figure~\ref{full-kris}. Next, the team went through a process of correlation between the KRIs and evaluating the potential strength of the risk indicator to rank the indicators. Finally, the team assessed and prioritised each risk indicator based on its potential to be extracted from the SoA documents. This process resulted in 6 KRIs being identified for implementation as part of the pilot. Table~\ref{table-kris} summarises these 6 key risk indicators and their interpretation.

\begin{table*}
\caption{Summary of regulatory requirements and the Key Risk Indicators (KRIs) of non-compliant advice}\label{table-kris}
\begin{tabulary}{\textwidth}{LLL}
\hline
& Key Risk Indicator (KRI) & Regulatory requirements and risks to be detected\\
\hline
\hline
1 & Goal-advice mapping & Have all of the client's goals been addressed by appropriate recommendations? \\
2 & Asset class diversification & Does the advice recommend a diversified investment strategy? \\
3 & Client's position after advice & Does the SoA consider position of the client post advice and does it include long-term (10 year) projection? \\
4 & Cash flow positivity & Is the client cash flow positive (short-term position) after implementation of advice? \\
5 & Starting superannuation fund balance & Does the SoA indicate a low starting balance (e.g. less than $\$200,000$) for self-managed superannuation fund? \\
6 & Insurance consideration & Does the SoA recommend life, trauma, TPD (Total \& Permanent Disability) and income protection insurance? \\
\hline
\end{tabulary}
\end{table*}

\section{KRIs Modeling Approach} \label{sec-models}

The team developed a number of machine learning, deep learning and rules based models to extract and score the 6 KRIs (described in Table~\ref{table-kris}) in the SoA documents, and validated these models on a diverse collection of 200 SoA documents from various financial institutions supplied by the regulatory agency. To cater for strict access control and privacy concerns regarding the handling of data with Personally Identifiable Information (PII), all handling and processing of SoAs data was performed within the regulatory agency's secure environment.

We briefly describe our approach for extracting and scoring each of the KRIs below.

\subsubsection{\textit{1. Goal-advice mapping:}}
As described in the earlier section, an SoA document is a legal contract between the customer and the financial advisor. It lists the customer's goals for seeking advice, the advisor's recommendations and the reasoning behind the recommendations. One of the significant tests from a regulatory compliance point of view is completeness of advice - that is, whether the advisor has addressed all of the citizen's goals. To extract this indicator, we developed a Long Short-Term
Memory (LSTM) classifier that is able to identify all sentences in the SoA document as either a goal, a recommendation or neither. We then developed an entailment model which maps all identified goals to identified recommendations along with a confidence rating for the mapping~\cite{Chen2019}. Any goals that are found to have no mapped recommendations are deemed to be not addressed and the risk indicator would be marked as high risk. Similarly, any goals that are mapped to recommendations with a low confidence score are similarly marked as medium risk. Finally, all goals that are mapped to recommendations with a high confidence score are marked as low risk.

\subsubsection{\textit{2. Asset class diversification:}}

In this KRI, the regulator's intent is to determine whether the advisor has recommended a diversified investment strategy for the client. The agency has defined its own classes of assets for the different types of investments. Diversified investment recommendations are from two or more of these asset classes. Diversified investment is generally low risk to the client and in the client's best interest with respect to minimising investment risk. In the majority of cases, investment information is found in tables and therefore, we applied a combination of technologies to extract this risk indicator. We extracted a large set of features such as number of monetary amounts, number of year series, number of positive/negative numbers and bigrams to train a random forest (RF) classifier to identify whether a table is an asset class table, projections table or cashflow table. We then applied a rules based model on the asset class tables to determine whether the asset class table contains a diversified allocation of asset classes.

\subsubsection{\textit{3. Client's position after advice:}} From the regulator's perspective, this  KRI aims to test whether the advisor has considered long term impact of the advice on the client's financial position. From past experience, the reviewers had found that in many cases the advisors had not considered how the advice impacts the client's financial situation. In other cases, the advisors had made statements such as "on implementation of the recommendations, we anticipate that your cash flow will improve by -X". In this case, the advisor had given consideration to the impact of advice but failed to acknowledge that the client will be in a worse position. This would be a violation of the best interest duty and marked as high risk. However, if the SoA contains statements such as  "on implementation of the recommendations, you will have a reduced cash flow but you will be able to achieve your investment goals", then this is a low risk case as the advisor has clearly considered the client's position after advice and explained the negative impact on cash flow and why this recommendation is good for the client overall. Similar to the KRI 2 above, we applied a combination of approaches to extract this KRI. We trained a Convolutional Neural Network (CNN) classifier to identify sentences in the SoA that discuss the client's capital/financial position. We then applied sentiment analysis on these statements to determine whether the advisor considers the position to be positive or negative. We augmented this with the check for capital projections table using the random forest table classifier described earlier. Finally, we extract numerical quantities and determine whether they are negative or positive in both the position statements and projections table to determine the risk rating of this KRI.

\subsubsection{\textit{4. Cash flow positivity:}}
This KRI is related to KRI 3 above. The intent is to test whether the advisor has considered the impact of the recommendations on the client's cash flow. Absence of cash flow analysis indicates a high risk of violation of ``best interests" duty. Similarly, analysis that shows negative cash flow for the client without statements why this is still a good recommendation is another indicator of high non-compliance risk. To extract this KRI, we applied the RF table classifier described earlier to determine whether the SoA contains post advice cash flow analysis. We then applied a rules based model to determine whether this analysis shows a positive outcome for the client.

\subsubsection{\textit{5. Starting superannuation fund balance:}}
For personal financial advice related to Self Managed Superannuation Funds (SMSFs), one of the key factors which determines whether the advice to switch to an SMSF is likely to be in best interests of the client is the starting fund balance. When it is difficult to determine the starting superannuation fund balance information, or when the balance is lower than a threshold (close to $\$200,000$), an SMSF is likely to be a poor recommendation for the client. Often the balance information is present in tables and multiple tables would have multiple balance amounts (with fees considered/excluded, balance for various funds listed separately etc.). For this reason, the auditors preferred to have all balance information extracted and presented to them rather than a binary indicator of whether the starting balance is less than or greater than a threshold. Therefore, we applied a combination approach to extract this KRI. We trained a CNN classifier to identify data units (sentences and tables) that discuss starting super balance. We then applied rules to extract the balance amounts and presented various statistics such as mean and median balance amounts for each data unit as well as aggregated for the whole SoA. This allowed the auditors the flexibility of using this information in conjunction with other client circumstances to make a subjective decision on whether the SoA meets this KRI.

\subsubsection{\textit{6. Insurance consideration:}} This KRI aims to test whether the advisor has considered risk minimisation for the client with respect to unforeseen circumstances negatively impacting their financial situation. Discussions with auditors determined that typically three different categories of insurance discussions are present in SoA documents. Risk rating for scoping out insurance discussion is subject to client's circumstances. Deferring insurance discussion to later is medium risk. Not considering insurance at all is high risk. Consequently, we trained a CNN classifier to identify sentences in the SoA that recommend personal insurance, defer insurance discussion and scope out insurance discussion from the SoA. We aggregated the results at the SoA level then applied a rules based model to determine whether the SoA satisfies this KRI.

\section{Outcomes of the Pilot}

\subsubsection{Immediate outcomes for the regulatory agency:}

All of the AI models built during the pilot were supervised models. Each model was evaluated on a hold out set of data. Once all the models were complete, a separate set of SoA documents was processed through all the models to generate a traffic light risk rating per unit of data (either a sentence or a table). All the risk ratings were then aggregated and presented per SoA document for the analysts and auditors to review. For the purposes of the pilot, final results were delivered to the analysts/auditors in a spreadsheet which highlighted the overall risk rating of each SoA, along with a breakdown of the individual KRIs within each SoA. The spreadsheet format provided the analysts with a familiar tool in which they could further rank, slice and dice the results to drive more detailed investigations.

The pilot delivered outcomes in two categories for the government agency:
\begin{itemize}
\item{Knowledge}:
	\begin{itemize}
	\item Understanding of AI (NLP, ML and DL) techniques for identifying Key Risk Indicators (KRIs) in Statement of Advice (SoA) documents
	\item Application and evaluation of a set of AI techniques for evaluating SoA KRIs
	\item A repeatable process for selecting and applying AI techniques on the remaining set of KRIs
	\end{itemize}
\item{Technical Assets}:
A set of AI models that can be applied to new SoA documents to assess them against a key set of KRIs.

\end{itemize}

Based on the combination of risk rating and KRI type, the overall risk rating for an SoA could be different. For example, if all KRIs are deemed to be equally important, the risk-rating for an SoA would rely on aggregating the KRIs to calculate an overall risk score for the document. However, not all KRIs are created equally and some KRIs are deemed of high significance; that is, if that KRI's risk is red, then the SoA document should also have a high risk rating regardless of the rating of the other KRIs. Furthermore, the importance/impact of each KRI could vary based on the customer's circumstances. Therefore, our approach allows for the flexible interpretation of risk and supports discretionary judgement from the auditors.

The pilot demonstrated the feasibility of automated review of a large number of SoAs with key risk indicators. The system can be used to refine the agency's targeting of client file reviews to the ones that are flagged as high risk by the system, thereby, speeding up the efficiency of the ``client file reviews" process.

\subsubsection{Broad reach outcomes:}

\begin{itemize}

\item Our system is designed to be used by both government auditors and advisors which will be described in detail in Section ~\ref{sec-application}. Therefore, this pilot also demonstrated the potential power of `sharing AI' between the public and private sectors. By sharing the same AI models with financial advisors, the advisors would be able to validate their advice against government regulations \textit{before} the advice is delivered to the client. This would greatly increase the quality of financial advice delivered to clients; a benefit for the government and citizens alike.

\item Another key outcome for the government regulation agency was that the experience of completing this pilot paved a pathway for creating a symposium inviting private sector, research institutions and financial institutions alike to discuss how regtech can be used to improve compliance in Australia's financial industry and advice outcomes for consumers. Furthermore, the symposium invited demonstrators to develop and showcase methodologies that can help determine features and characteristics of the financial advice documents in the client files. This strongly indicates that the experience and insights obtained in the pilot were pivotal in the conceptual design of this symposium. Therefore, this pilot has become a stellar example of using AI to encourage public service innovation.

\end{itemize}

\section{Challenges and Lessons Learned}

A major challenge in applying AI models for regulatory compliance is on the mapping of a relevant regulation to a risk indicator and devising a methodology to identify that indicator from an unstructured data source such as an SoA document. Similar to other AI projects conducted by our lab, we scheduled a number of intense workshops of 1-2 days each at the start of this project in which Researchers and SMEs (in this case auditors and analysts) collaborated to define the scope of the project. This initial process required close collaboration between SMEs who in this case identified and described the process they use to evaluate the indicators and explained the interrelationships between the different indicators; and Researchers who applied AI techniques to automate evaluation of a set of indicators. The success of the pilot was defined by the strength of the collaboration between Researchers and SMEs with each bringing their deep expertise to address this challenge.

A further challenge is subjectivity in the interpretation of a regulation. Often reviewers apply subjective judgement on the interpretation of a regulation based on previous findings and personal experience. For example, past experience indicates that for a client whose starting super balance is lower than \$200,000, a Self Managed Super Fund (SMSF) is likely to be a bad recommendation. However, when the balance amount is between \$200,000 and \$250,000 there is room for subjectivity in the interpretation and the risk status may vary based on the reviewer's experience and other client details. The challenge is then what changes need to be made to the modelling approach such that there is opportunity for subjectivity for the decision makers.

Another example is when a client specifically states that they want to switch to a Self Managed Super Fund (SMSF). Some reviewers were okay with this being a good enough reason for the recommendation (regardless of the client's circumstances) whereas others thought that the client's circumstances should be given more consideration by the advisor than their explicit wishes.

For successful AI projects it is important to understand the limitations of current technology. Broad reasoning involving unbiased human subjectivity is not an area which AI can address to a sufficiently high degree of accuracy that is necessary to meet the high standards required of government agencies. This is one of the reasons our solution is specifically designed to aid auditors and SMEs in making their final decision; the solution is not designed to make the actual decision itself as human subjectivity is a vital piece in the decision making process.

\begin{figure*}[hbt!]
\centering
{\includegraphics[width=0.8\textwidth]{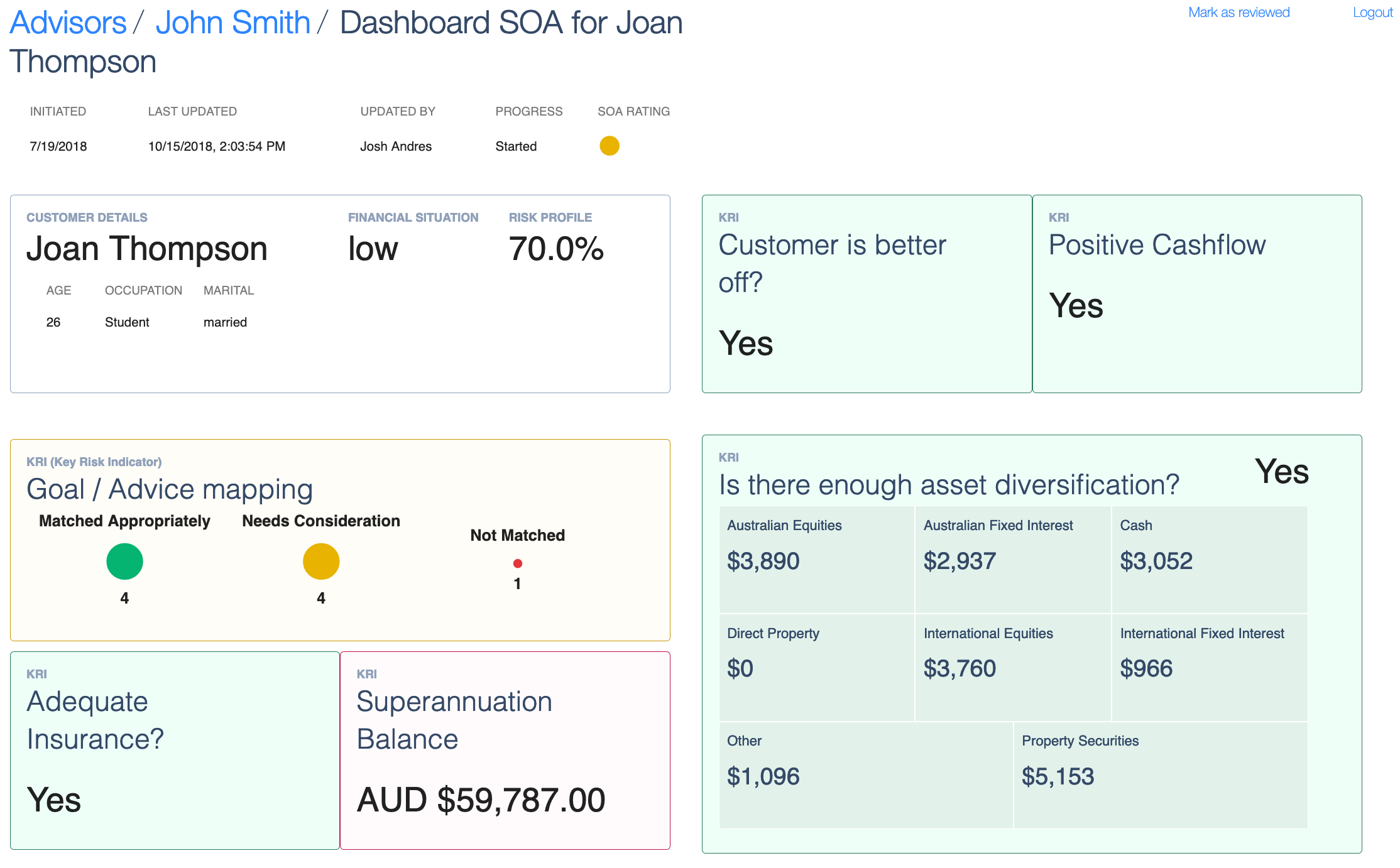}}
\caption{Overview of an SoA with traffic light rating of the Key Risk Indicators of non-compliance.}
\label{ui-soa}
\end{figure*}

Another challenge we encountered was how to interpret a regulation such that it could be modelled using available AI techniques. One of the KRIs is focused on insurance which was initially interpreted as: whether the advisor has considered adequate personal insurance for the client. Because of its subjectivity, adequacy is difficult to quantify using an AI model. Therefore, the team modified the test to capture the risk using slightly different criteria. For example, on further discussion with the reviewers and detailed examination of sample SoAs, we discovered four different types of insurance discussions in the SoAs. In some SoAs, insurance discussion is scoped out of the SoA .
In others, the SoA defers insurance discussion to a later date.
Still further, other SoAs recommend various personal insurance products as either new products or replacement for the client's existing products. Finally, the most problematic SoAs are the ones that do not include insurance considerations for the client.
In discussions with the auditors, this risk test was modified from
 "has the advisor considered adequate personal insurance for the client" to "what kind of insurance consideration does the SoA provide among the four different options- scoped out, deferred, recommended, none". Based on the evaluation of an SoA on these four categories, the auditors now have the discretionary power to apply their judgement on whether this KRI is met by the SoA. Similar to the cases discussed before, in this case as well there was difference of judgement among the auditors on the risk rating.
By providing granular risk information, our approach to interpretation of the relevant regulation offer the reviewers the clarity and confidence to make their decisions.

\section{Application Workflow}\label{sec-application}

On completion of the pilot, we deployed the KRI models described in Section~\ref{sec-models} in an application that can be used by both advisors and auditors/reviewers. Each SoA document is ingested and parsed by the system, then a pipeline of KRI models is applied to them. The result is traffic light annotation of the document for each KRI which is then presented in the user interface. Figure~\ref{ui-soa} gives an overview of an SoA as presented by our system with traffic light panels for each KRI.

\begin{figure*}[hbt!]
\centering
{\includegraphics[width=0.8\textwidth]{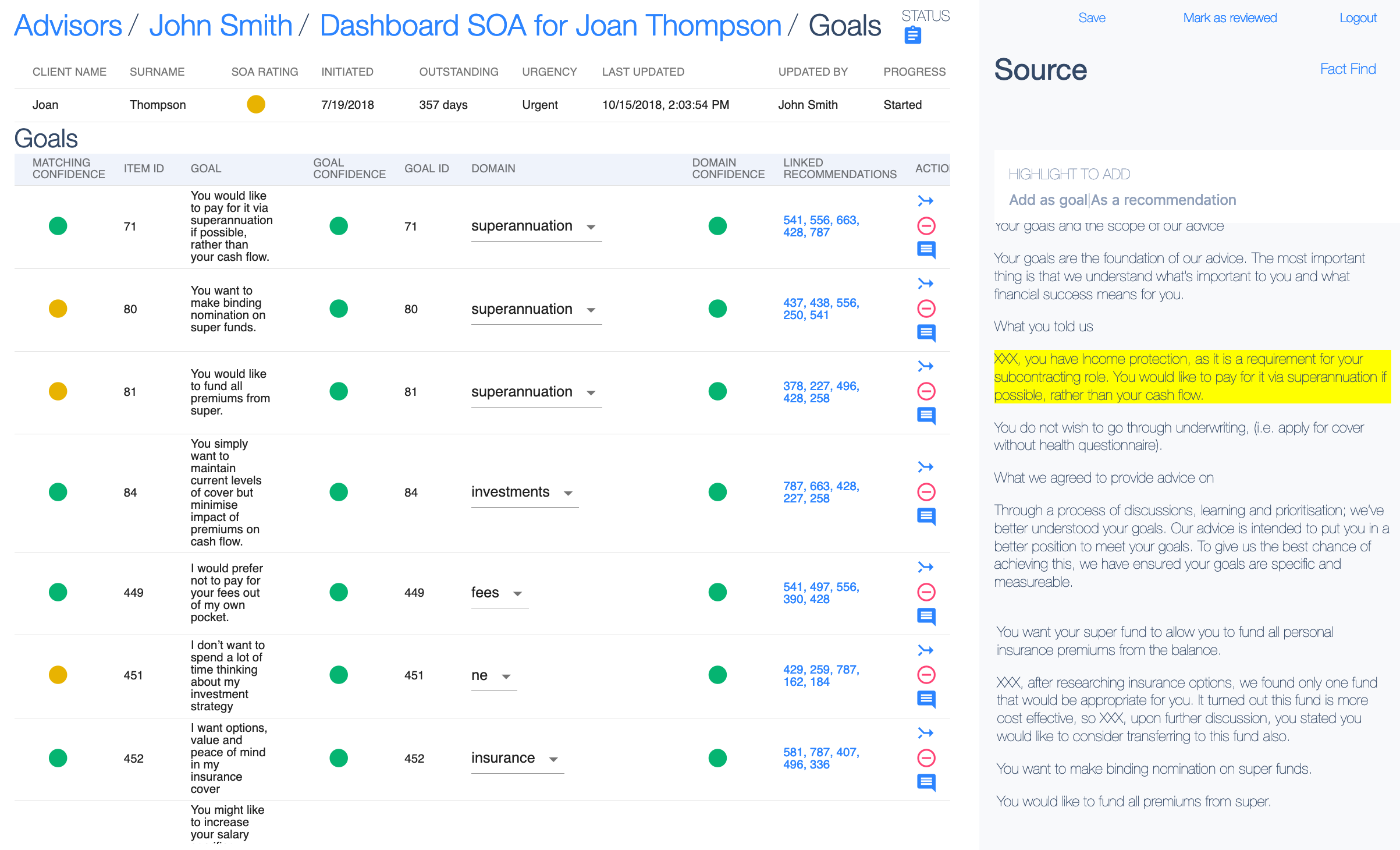}}
\caption{Drill down view of goal-advice mapping with original SoA document on the right.}
\label{ui-goals}
\end{figure*}

The user can drill down on each KRI panel to view the details behind the rating for the KRI. Figure~\ref{ui-goals} shows the drill down for goal/advice mapping panel. The left panel shows all goals identified by the system. Each goal has a set of linked recommendations mapped by the system, the strength of mapping is depicted by the traffic light rating under the column \begin{scriptsize}MATCHING RTNG\end{scriptsize}. On the right panel the original SoA document is loaded for reference. The user is able to merge goals, delete goals and add new goals by highlighting text in the right panel. The user can perform similar actions with recommendations and with other KRIs. Further, the application offers the user the capability to add comments explaining the reasons for the updates. All of these user actions are captured by the system for auditing purposes and used for further refining the AI models.

A single application for use by both advisors and auditors ensures that transparency is maintained in the audit process and the auditors have access to the extra/supplementary information from the advisors in the form of comments. Our solution provides a regulatory compliance auditor view and an advisor view, allowing feedback made by both parties to be recorded and be visible for review. Continued use of such a system will ensure that all personal financial advice is compliant before it is delivered. This will lead to improved financial well-being for citizens and as a knock-on will reduce the workload on auditors who can focus on higher-value tasks.

Furthermore, our solution overcomes obstacles in compliance regulation by facilitating the analysis of advice documents in bulk (taking a little over two minutes to analyse a single document, instead of the hours typically taken to manually review an advice document).
This, in turn, promotes greater vigilance of compliance, and thus increases the transparency of the review process at minimal time and cost to both regulatory agencies and the regulated individuals.

\section{Discussion}

Building successful Artificial Intelligence systems such as the one described herein requires the melding of two different skills sets; the deep skills required to build the AI models, along with deep domain knowledge of Subject Matter Experts (SMEs) in the field the system will operate.
A subset of the KRIs discussed in Section 3 were developed in partnership between IBM Research,  an Australian government regulatory body and Promontory Financial Group, a private consultancy firm specialising in regulatory compliance. In this partnership, IBM Research brought the know-how and skills in Artificial Intelligence (natural language processing, machine learning and deep learning) required to extract salient information from complex financial documents, along with the ability to build, test and validate Artificial Intelligence models capable of mimicking a set of cognitive functions of the auditors. The regulation agency and the consultancy firm brought the SMEs, the analysts, the auditors, who have the deep domain knowledge to guide which KRIs to build, and how each of those KRIs should be assessed for compliance.
To enable skills transfer across into the public sector several employees from the regulatory agency co-located with IBM Research during the course of the pilot, providing an insight into the skills, tools and techniques required to build AI models. Until AI becomes mainstream within government IT departments, fostering this type of relationship between private and public sectors will be key to the early advancement of AI within government departments.

Adoption of AI models in government internal processes can be further supported by ensuring that the results presented by the models are in a format that the users are familiar with, can easily consume and aid the SME's decision process. This will promote acceptance and adoption of the models internally which is often the first hurdle in infusing AI in government and public sector.

Furthermore, successful introduction of AI models into internal processes of government departments requires that the AI models being introduced be designed to improve the efficiency of the people doing the job, not replace them. By their very nature, AI systems deployed into the government sector will be held to a higher standard, and must act in the interest of the public. AI technology has much promise, but it is still in its infancy and as such the technology should be deployed with human over-sight in situations in which its outcomes directly impact citizens.
As impressive as our system is, it cannot do the job of the auditor or the financial advisor. It can only aid them and help them become more efficient.

\section{Summary}

In this paper we have described a pilot with an Australian Government regulation agency conducted between July 2018 to September 2018. The pilot focused on (i) defining a set of Key Risk Indicators (KRIs) that map applicable regulation and advisor conduct guidelines to features that can be extracted in Statement of Advice (SoA) documents, and (ii) implementing a set of Artificial Intelligence models and rules based models to extract these features and provide indication of regulatory compliance status of an advice document. Our solution reduces the time taken for auditing advice documents from several hours to minutes and enables bulk analysis of advice documents; highlighting documents that are most at-risk along with the sources of those risks. This makes our solution a critical tool for assessing regulatory compliance of financial advice documents.

The solution described in this paper aims to assist rather than replace the analysts and auditors who make compliance judgements. Since this type of solution clearly benefits the government and its citizens whilst retaining human subjectivity in the decision process, it is one approach to gradually introducing and encouraging the adoption of AI within the public sector.
From this experience of working together, the government regulatory agency has since created a symposium inviting private sector, research institutions, financial institutions alike to come up with ideas and solutions applying AI in regulatory compliance. In this respect, this pilot has become an exemplar of using AI to encourage public service innovation. Finally, the pilot demonstrated how a public-private partnership consisting of researchers from a private organization with deep expertise in AI technologies and analysts and auditors from a government agency with deep subject matter expertise can effectively collaborate to solve seemingly intractable problems such as extracting and rating nuanced risk factors in unstructured advice documents.

\bibliography{ai-for-government-pfa}
\bibliographystyle{aaai}
\end{document}